\documentclass[12pt]{iopart}

\usepackage{graphicx}  
\usepackage{epsfig}

\begin{document}

\title[]{Statistical Methods for Thermonuclear Reaction Rates and Nucleosynthesis Simulations}

\author{Christian Iliadis}
\address{University of North Carolina at Chapel Hill, Chapel Hill, NC 27599-3255, USA; and
Triangle Universities Nuclear Laboratory, Durham, North Carolina 27708-0308, USA}
\ead{iliadis@unc.edu}

\author{Richard Longland}
\address{North Carolina State University, Raleigh, NC 27695, USA; and
Triangle Universities Nuclear Laboratory, Durham, North Carolina 27708-0308, USA}

\author{Alain Coc}
\address{Centre de Sciences Nucl\'eaires et de Sciences de la
Mati\`ere  (CSNSM), CNRS/IN2P3, \\ Universit\'e~Paris~Sud~11, UMR~8609,
B\^atiment 104, F--91405 Orsay Campus, France}

\author{F. X. Timmes}
\address{School of Earth and Space Exploration, Arizona State University, Tempe, AZ 85287-1504, USA}

\author{Art E. Champagne}
\address{University of North Carolina at Chapel Hill, Chapel Hill, NC 27599-3255, USA; and
Triangle Universities Nuclear Laboratory, Durham, North Carolina 27708-0308, USA}

\begin{abstract}
Rigorous statistical methods for estimating thermonuclear reaction
rates and nucleosynthesis are becoming increasingly established in
nuclear astrophysics. The main challenge being faced is that
experimental reaction rates are highly complex quantities derived from
a multitude of different measured nuclear parameters (e.g.,
astrophysical S-factors, resonance energies and strengths, particle
and $\gamma$-ray partial widths). We discuss the application of the
Monte Carlo method to two distinct, but related, questions. First,
given a set of measured nuclear parameters, how can one best estimate
the resulting thermonuclear reaction rates and associated
uncertainties? Second, given a set of appropriate reaction rates, how
can one best estimate the abundances from nucleosynthesis (i.e.,
reaction network) calculations? The techniques described here provide
probability density functions that can be used to derive statistically
meaningful reaction rates and final abundances for any desired
coverage probability. Examples are given for applications to s-process
neutron sources, core-collapse supernovae, classical novae, and big
bang nucleosynthesis.
\end{abstract}

\maketitle

\section{Introduction}\label{sec:Intro}
Our understanding of the universe has been revolutionized by new
observational technologies. Advances in detectors, computer processing
power, network bandwidth, and data storage capability have enabled new
sky surveys (e.g., the {\it Sloan Digital Sky Survey}
\cite{abazajian_2009_aa}).  Advances have triggered many new optical
transient surveys (e.g., the {\it Palomar Transient Factory}
\cite{rau_2009_aa,law_2009_aa}) that probe ever larger areas of the
sky and ever-fainter sources, opening up the vast discovery space of
time-domain astronomy. Advances have also allowed for space missions,
for example, NASA's {\it Kepler} \cite{koch_2010_aa,batalha_2010_aa},
that continuously monitor more than 100,000 stars. The discoveries
from these surveys include revelations about stellar nucleosynthesis,
unusual explosion outcomes, and remarkably complex binary star
systems. The immediate future holds tremendous promise, as both the
space-based survey {\it Gaia} and the ground based {\it Large Synoptic Survey
Telescope} come to fruition.

Many forefront questions in astrophysics ultimately require a detailed
quantitative knowledge of stellar properties, thus challenging stellar
models to become more sophisticated, quantitative and realistic in
their predictive power (e.g., {\it Modules for Experiments in Stellar Astrophysics}
\cite{paxton_2011_aa,paxton_2013_aa}).  This in turn requires more
detailed physics input, such as thermonuclear reaction rates and
opacities, and a concerted effort to validate models through
systematic observations. The study of nuclear reactions in the
observable universe remains at the forefront of nuclear physics and
astrophysics research. On the nuclear physics side, data on cross
sections and nuclear properties of astrophysically important reactions
are being obtained at radioactive ion-beam facilities and at
stable-beam facilities at an accelerated pace (e.g., {\it Facility for
 Rare Isotope Beams} \cite{balantekin_2014_aa}).  Radiation
detectors, ion beam technology, and low-background techniques have
reached an unprecedented stage of sophistication, permitting
measurements of increasing precision and sensitivity (e.g., {\it
Laboratory for Experimental Nuclear Astrophysics}
\cite{cesaratto_2010_aa,longland_2006_aa}).  We recognize that insight
can be gained in observational astronomy and nuclear astrophysics by
acquiring new information about atomic nuclei and by recognizing the
importance of what needs to be measured in the laboratory.

Thermonuclear reaction rates are at the heart of every big bang and
stellar model. In the following we will discuss statistical techniques
for deriving reliable reaction rates and their associated
uncertainties. This approach is useful for assessing which nuclear
properties of a given reaction need to be measured in the laboratory,
and for identifying the most important nuclear reactions that impact a
given isotopic abundance during nuclear burning in stars.

Experimental thermonuclear reaction rates, based on nuclear physics
input gathered from laboratory measurements, were first presented by
Willy Fowler and collaborators more than 40 years ago
(Ref.~\cite{caughlan_1988_aa}, and references therein). Those reaction
rates were directly based on nuclear physics experiments and were
distinct from reaction rates derived from theory (e.g., the
Hauser-Feshbach model). The incorporation of Fowler's rates into
stellar models represented a paradigm shift for astrophysics.  With a
solid nuclear physics foundation, stellar simulations could provide
reasonable estimates of nuclear energy generation and
nucleosynthesis. Subsequent work \cite{angulo_1999_aa,iliadis_2001_aa}
incorporated newly measured nuclear physics data, but the reaction
rates were still computed using techniques developed prior to 1988.

The main challenge with such approaches in the modern era, is that
thermonuclear reaction rates are reported as single value at a given
temperature, without any uncertainty estimate, or that a recommended
rate is presented together with ``limits''.  These upper and lower
rate limits are frequently obtained by inclusion or exclusion
of unobserved low-energy resonances. Such reaction rate limits are a
drawback in the modern era since they lack a rigorous statistical
meaning. Specifically, since the reaction rate probability density
function remains unknown with these methods, the reported rate limits
cannot be quantified in terms of a coverage probability. A significant
obstacle to overcome in this regard is the fact that thermonuclear
reaction rates are highly complex quantities derived from a multitude
of nuclear physics properties painstakingly extracted from laboratory
measurements (resonance energies and strengths, non-resonant cross
sections, spectroscopic factors, etc.).

\section{Monte-Carlo Based Reaction Rates}\label{sec:MCRates}
\subsection{Method}\label{sec:newmethod}

One approach to addressing these challenges is described in
Refs.~\cite{Smith:2002uk,longland_2010_aa,iliadis_2010_aa,
  iliadis_2010_ab,iliadis_2010_ac}. The method is conceptually
straightforward and follows a Monte Carlo procedure. First, all of the
measured nuclear physics (input) properties entering into the reaction
rate calculation are randomly sampled according to their individual
probability density functions. Second, the sampling is repeated many
times and thus provides the Monte Carlo reaction rate (output)
probability density. Third, the associated cumulative distribution is
determined and is used to derive reaction rates and their
uncertainties with a precise statistical meaning (i.e., a quantifiable
coverage probability).  For example, for a coverage probability of
68\%, the low, recommended, and high Monte Carlo rates can be defined
as the 16th, 50th, and 84th percentiles, respectively, of the
cumulative reaction rate distribution\footnote{$N$ sampled values of reaction rates, $x_i$, are
  sorted into ascending order and the percentile, $q$, is found from
  the fraction of values located below a given value of $x_q$.}. The
main challenge is to randomly sample all nuclear physics input
parameters, including resonance energies and strengths, partial
widths, reduced widths, astrophysical S-factors, etc., according to
physically motivated probability density functions
\cite{longland_2010_aa}.

Depending on the nature of the nuclear physics observable, the (input)
probability densities should be chosen according to the central limit
theorem of statistics. It states that the sum of $n$ independent
continuous random variables $x_i$, with means $\mu_i$ and standard
deviations $\sigma_i$, becomes a normal (Gaussian) random
variable in the limit of $n \rightarrow \infty$, independent of the
form of the individual probability density functions of the
$x_i$. Many measurement uncertainties are treated as Gaussian random
variables if it can be assumed that the total uncertainty is given by
the sum of a large number of small contributions. This is
usually the case for measured resonance energies, with contributions
from the beam energy calibration, the measured yields, the fitting of
the yield curve to find the 50\% point, target inhomogeneities, dead
layers, etc.

It also follows directly from the central limit theorem that a random
variable will be distributed according to a lognormal density
function if it is determined by the product of many
factors. This is usually the case for experimental resonance strengths
(i.e., integrated resonance cross sections), which are determined from
the measured number of counts of a thick-target yield, the integrated
beam charge, a detector efficiency, a stopping power, etc. The
lognormal distribution is given by
\begin{equation}
f(x) = \frac{1}{\sigma \sqrt{2\pi}} \frac{1}{x} e^{-(\ln x - \mu)^2/(2\sigma^2)}  \label{lognormalpdf}
\end{equation}
and is defined by the two parameters $\mu$ and $\sigma$. The first
parameter $\mu$ determines the location of the distribution, while the
second parameter $\sigma$ controls the width. An
exhaustive account of this method can be found in
Ref.~\cite{longland_2010_aa}.

\begin{figure}
\begin{center}
\includegraphics[width=3.5in]{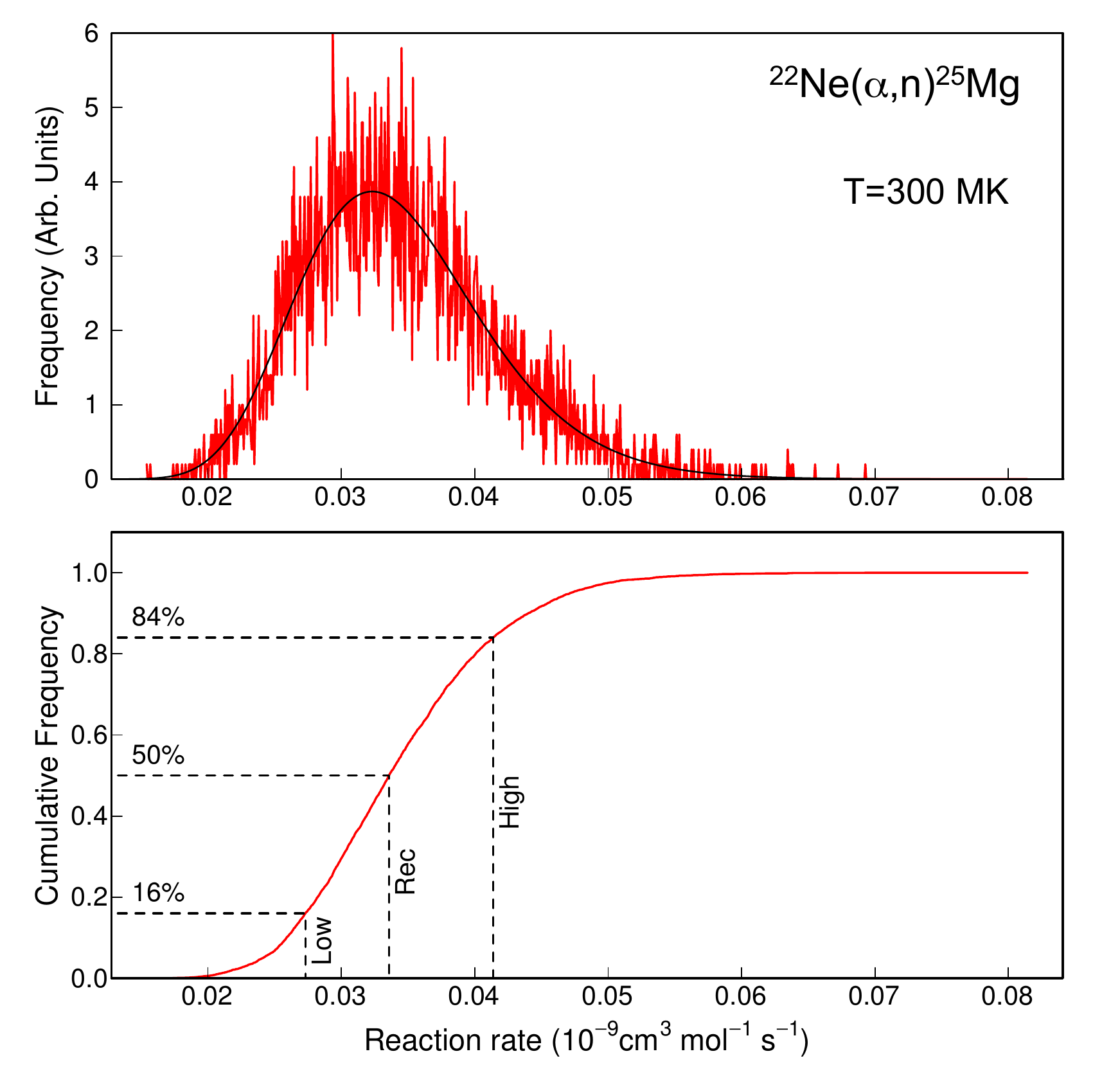}
\caption{Experimental Monte Carlo based reaction rates for the crucial
$^{22}$Ne($\alpha$,n)$^{25}$Mg neutron source in the s-process at a
temperature of 300 MK \cite{Longland:2012ix}.  The rate is obtained by
sampling 166 different nuclear parameters (resonance energies and
strengths, $\alpha$-particle, neutron, and $\gamma$-ray partial
widths), and the Monte Carlo sampling is repeated 10,000 times. (Top)
Reaction rate probability density function, shown in red; the black
solid line represents a lognormal approximation, which is directly
obtained from the mean and variance of the Monte Carlo rate samples
(i.e., no fitting is involved). (Bottom) Cumulative reaction rate
distribution; notice the much reduced scatter. The vertical dotted
lines represent the low, median and high Monte Carlo reaction rates,
which are obtained from the 16th, 50th and 84th percentiles,
respectively. From A.~E. Champagne, C. Iliadis, and R. Longland, AIP Advances 4, 041006 (2014).}
\label{fig:ne22an}
\end{center}
\end{figure}

As an example, we show in Fig.~\ref{fig:ne22an} the experimental Monte
Carlo based rate of the $^{22}$Ne($\alpha$,n)$^{25}$Mg reaction at a
temperature of 300 MK, which is a key neutron source for the
astrophysical s-process that occurs during helium burning in AGB stars and massive
stars. The total reaction rate has contributions from
23 resonances with measured energies and resonance strengths or
partial widths, and from 19 resonances for which only upper limits on
the partial widths are available. In total, 166 different nuclear
parameters (resonance energies and strengths; $\alpha$-particle,
neutron, and $\gamma$-ray partial widths) are randomly sampled to
estimate this particular rate. The upper and lower panels of the
figure display the Monte Carlo probability density function and the
associated cumulative distribution, respectively, of the 
{\it experimental} rate at a stellar temperature of 300 MK. The 16th,
50th, and 84th percentiles, indicated in the lower part, define one
particular choice for the reaction rate uncertainty, corresponding to
a coverage probability of 68\%. The Monte Carlo based rate of this
reaction differs significantly from previously published
results. Details can be found in Ref.~\cite{Longland:2012ix}.  The
Monte Carlo probability density is a smoothly varying function,
without any sharp boundaries, challenging previous definitions of an
``upper limit'' or ``lower limit" for a reaction rate.

Reaction rates based on the Monte Carlo
method do not consider only {\it statistical} uncertainties in the nuclear
physics input, since the Monte
Carlo sampling does not distinguish between statistical and systematic
effects. For example, suppose a reported experimental value for a
resonance strength amounts to 5.0$\pm$0.5~eV. Closer scrutiny of the
experimental technique may reveal that certain systematic effects were
not taken into account. Perhaps the detection efficiency was not
corrected for coincidence summing effects, or older stopping power
values were employed. If such systematic effects can be accounted for
by adjusting the reported mean value and the uncertainty, the nuclear
input to the Monte Carlo procedure will represent the {\it best
estimate} based on known statistical and systematic effects.  Of
course, unknown systematic effects may still impact the total rate.

A public web-portal interface to a RatesMC executable \cite{longland_2010_aa}\footnote{See \texttt{http://starlib.physics.unc.edu}.} allows users to calculate experimental Monte Carlo based rates. 
The current version of the code is applicable to many nuclear reactions provided that the total rates are determined by the incoherent contributions of any number of broad or narrow resonances and of up to two non-resonant (i.e., direct) amplitudes. The code also accounts for interferences between any two resonant amplitudes. Interferences of more than two resonant contributions, or between a resonant and a non-resonant amplitude, have not been implemented yet.

Many correlations between nuclear quantities are considered carefully in RatesMC. For example, if the strength of a narrow resonance is estimated from a reduced width or a spectroscopic factor, then the uncertainty in the resonance energy enters both in the Boltzmann factor and in the penetration factor. Thus the same random value of the resonance energy, drawn from a Gaussian probability density function, must be used in both expressions. Other correlations are not yet considered in RatesMC, although their implementation is straightforward from a computational point of view. For example, the same partial width value may enter in the rate expressions of two competing reactions involving the same target nucleus. In this case, the same random value of the width, drawn from a lognormal probability density function, should be used for both reactions. We will consider such improvements in a future release of RatesMC.  

It is apparent that in example of Fig.~\ref{fig:ne22an} the rate probability
density can be approximated by a lognormal distribution, shown as the
black solid line in the top panel of Fig.~\ref{fig:ne22an}. We find
that this is the case for the majority of Monte Carlo based reaction
rates at most stellar temperatures of interest. A detailed discussion
is presented in Ref.~\cite{iliadis_2010_aa}. We will return to this
point below, which is crucial for how to implement the new Monte Carlo
based reaction rates in nucleosynthesis studies.

\subsection{Upper Limits of Nuclear Physics Input Parameters}
Many reaction rates have contributions from unobserved low-energy
resonances. More precisely, levels are known to exist near the
projectile threshold energy, but the corresponding resonances have not
been observed directly in the laboratory yet. At low bombarding
energies, proton or $\alpha$-particle partial widths are dominated in
magnitude by the transmission through the Coulomb barrier and thus are
usually much smaller compared to $\gamma$-ray partial widths. For
example, in the simple case of a low energy resonance with only one
particle channel and the $\gamma$-ray channel open, the resonance
strength that enters into the calculation of the reaction rate is
given by
\begin{equation}
{
   \omega\gamma\equiv\omega\frac{\Gamma_x \Gamma_\gamma}{\Gamma}\approx\omega\Gamma_x=2\frac{\omega\hbar^2}{\mu R^2}P_{\ell}\theta_x^2 
}
\label{eq:PT2}
\end{equation}
with $\Gamma_x$, $\Gamma_\gamma$, $\Gamma$ the particle partial width,
$\gamma$-ray partial width, and total width, respectively; $\mu$, $R$,
$P_{\ell}$, and $\theta_x^2$ denote the reduced mass, channel
(nuclear) radius, penetration factor, and dimensionless reduced width,
respectively; furthermore, $\omega\equiv(2J_r+1)/[(2j_p+1)(2j_t+1)]$,
where $J_r$, $j_t$, $j_p$ are the spins of the resonance, target, and
projectile, respectively. In simple terms, the penetration factor
represents the nuclear exterior and can be computed from numerical
values of Coulomb wave functions. The only unknown quantity in the
above expression, assuming that $J^{\pi}$ is known, is the
dimensionless reduced width, which describes the nuclear
interior\footnote{The dimensionless reduced width is closely related
  to the spectroscopic factor, see Ref.~\cite{Iliadis:1997dh}.}. The
question arises of how to implement such contributions into the Monte
Carlo sampling procedure if only an upper limit for $\theta_x^2$ is
available, either from experiment or from theory.

A solution to this problem is closely related to fundamental
predictions of random matrix theory. The basic assumption is that
energy levels in atomic nuclei at several MeV excitation energies
represent chaotic systems. The reduced width amplitude for formation
or decay of an excited compound nucleus is assumed to be a random
variable, with many small contributions from different parts of
configuration space. If the contributing nuclear matrix elements are
random in magnitude and sign, the reduced width amplitude is
represented by a Gaussian probability density centered at zero,
according to the central limit theorem
(Sec.~\ref{sec:newmethod}). Consequently, the corresponding reduced
width, i.e., the square of the amplitude, is described by a
chi-squared probability density with one degree of freedom,
\begin{equation}
  {
   g(\theta^2)=\frac{1}{\sqrt{2\pi \theta^2 \left< \theta^2 \right>}}e^{-\frac{\theta^2}{2\left< \theta^2 \right>}}
   }
\label{eq:PT3}
\end{equation}

with $\left< \theta^2 \right>$ the local mean value of the
dimensionless reduced width. This expressions is known as the
Porter-Thomas distribution \cite{porter_1956_aa}. It implies that the
reduced widths for a single reaction channel, i.e., for a given
nucleus and set of quantum numbers, vary by several orders of
magnitude, with a higher probability for smaller values of the reduced
width. Until recently this fundamental prediction of random matrix
theory had been disregarded in nuclear astrophysics. It was shown in
Refs.~\cite{iliadis_2010_aa,iliadis_2010_ab} that a proper treatment
of the contributions from unobserved resonances, based on the
Porter-Thomas distribution, can change the estimated total
thermonuclear reaction rate by orders of magnitude compared to
previous predictions.

The crucial ingredient for the Monte Carlo sampling of an upper limit
contribution according to the Porter-Thomas distribution is the mean
value of the reduced width, $\left< \theta^2 \right>$. It is not
predicted by random matrix theory, but can be obtained from the
analysis of laboratory data or from a suitable nuclear reaction
model\footnote{The mean reduced width is related to the strength
  function of channel $c$ via $s_c^J \equiv \left< \gamma^2_{\lambda
    c} \right>/D^J$, where $D^J$ is the mean energy spacing for
  compound levels of spin $J$; the reduced width, $\gamma^2$, and
  dimensionless reduced width, $\theta^2$, are related by $\gamma^2
  \equiv (\hbar^2/(\mu R^2))\theta^2$, with $\mu$ the reduced mass and
  $R$ the channel radius. The strength function determines the
  transmission coefficient, which is a key quantity for estimating
  average nuclear reaction cross sections.}. A first step in this
direction was the recent extraction \cite{pogrebnyak_2013_aa} of mean
reduced widths from high-resolution data measured at TUNL for target
mass ranges of $A=28-40$ ($\alpha$-particles) and $A=34-67$
(protons). For example, a mean value of $\left< \theta^2_\alpha
\right>$=0.018, averaged over target nuclei, spin-parities, and
excitation energies, was obtained for $\alpha$-particles, almost a
factor of two larger than the preliminary value suggested in
Ref.~\cite{longland_2010_aa}.

\begin{figure}
\begin{center}
\includegraphics[width=4in]{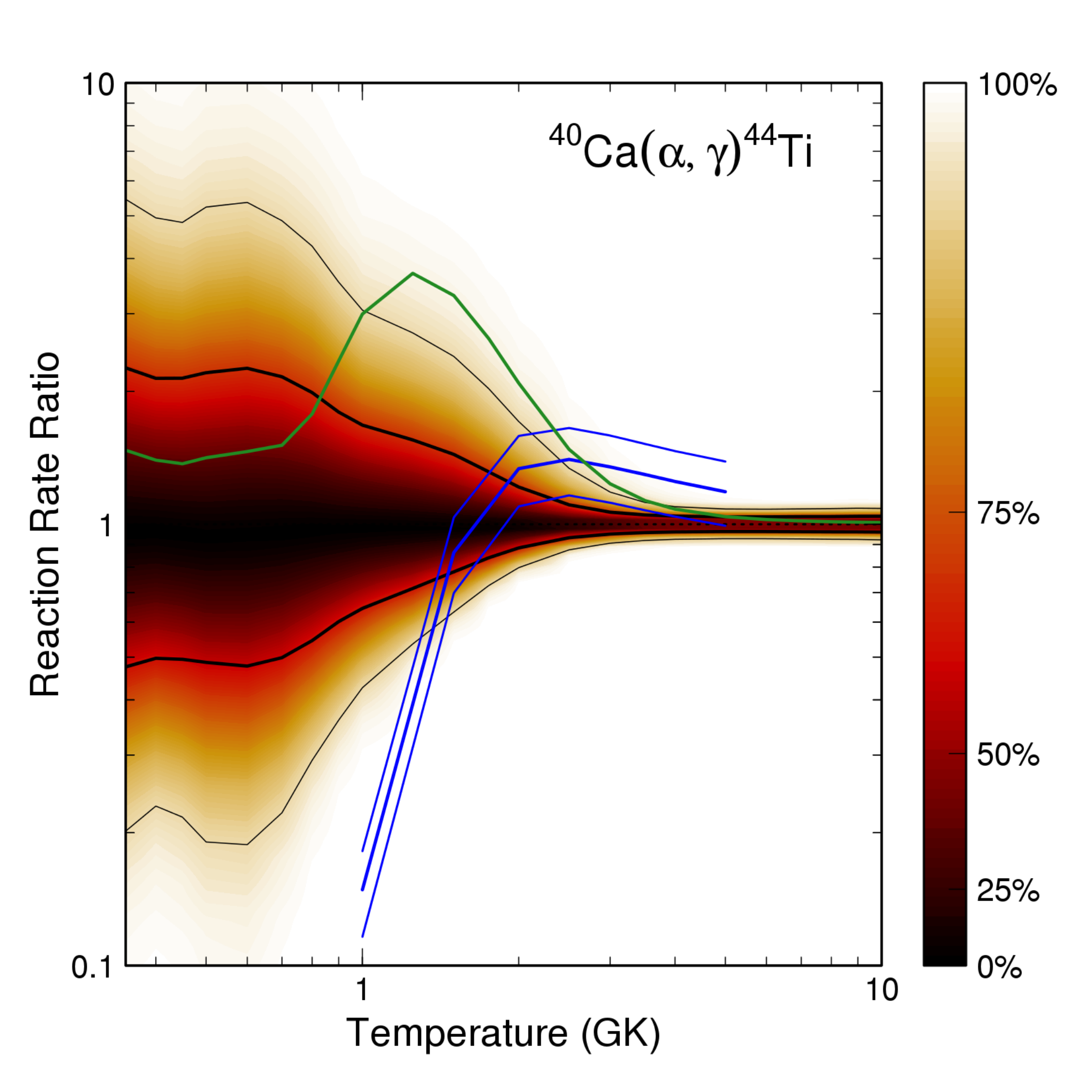}
\caption{Monte Carlo-based reaction rates of
$^{40}$Ca($\alpha$,$\gamma$)$^{44}$Ti. For a better comparison, the
rates are normalized to the recommended Monte Carlo rate. The
color-shading indicates the coverage probability in percent. The thick
(thin) black lines indicate the high (low) Monte Carlo rates for a
coverage probability of 68\% (95\%). Note that the Monte Carlo rate
has no sharp bounderies (i.e., no ``lower limit" or ``upper limit"),
but instead is represented by a smoothly varying probability density
function along the ordinate. The blue and green lines show the rates
obtained using conventional (i.e., pre-Monte Carlo) methods. From I. Pogrebnyak, C. Howard, C. Iliadis, R. Longland, and G. E. Mitchell, Phys. Rev. C 88, 015808 (2013), Copyright 2013 by the American Physical Society.}
\label{fig:ca40ag}
\end{center}
\end{figure}

An example for the relevance of these results is given in
Fig.~\ref{fig:ca40ag}. The calculated experimental Monte Carlo-based
rates for the $^{40}$Ca($\alpha$,$\gamma$)$^{44}$Ti reaction, which is
crucial for the production of the $\gamma$-ray emitter $^{44}$Ti in
core-collapse supernovae 
\cite{woosley_1973_aa,woosley_1992_aa,timmes_1996_ab,the_1998_aa,hoffman_2010_aa,magkotsios_2010_aa,larsson_2011_aa,grebenev_2012_aa}
are shown as a contour plot (black-red-yellow; see color bar on the
right). The colors signify the coverage probability between any given
rate boundaries. The rates are normalized to the recommended Monte
Carlo rate for a better comparison.  For example, the thick (thin)
black lines indicate the high and low Monte Carlo rates for a coverage
probability of 68\% (95\%). The blue and green lines show the rates
obtained by using conventional (i.e., pre-Monte Carlo) methods: (blue)
results of Ref.~\cite{robertson_2012_aa}, where the unobserved
low-energy resonances were disregarded; (green) upper limit obtained
if the maximum contribution of the unobserved resonance at
E$_{\alpha}^{c.m.}=2373$ keV is adopted. It is apparent that the new
Monte Carlo rates are significantly different from previous results.

So far, the only systematic analysis of mean values for dimensionless
reduced widths has been presented by Pogrebnyak et
al. \cite{pogrebnyak_2013_aa}.  As already mentioned, these values
were extracted from the available experimental data for a range of
compound nuclei, $A$, spin-parities, $J^{\pi}$, and excitation
energies, $E_x$. However, the experimental values cover only a small
part of the $A$-$J^{\pi}$-$E_x$ parameter space and it is highly
desirable to have access to $\left< \theta^2 \right>$ values for all
cases of interest. Considering that the data analyzed in
Ref.~\cite{pogrebnyak_2013_aa} were accumulated over a period of more
than 40 years at the now decommissioned high-resolution 3-MeV Van de
Graaff accelerator laboratory at Triangle Universities Nuclear
Laboratory, it is clear that the desired $\left< \theta^2 \right>$
values need to be obtained from nuclear theory, for example, using the
shell model. Additional efforts are needed in this direction.

\subsection{Individual Contributions to the Total Reaction Rate}
Suppose a given nuclear reaction has been identified as a key process
for some astrophysical environment and that an experimentalist intends
to measure this reaction. The immediate questions at hand are: what
energy range should be covered in the laboratory? And which nuclear
properties should, in fact, be measured? The usual approach is to
determine first the temperature range of astrophysical interest, then
to convert the temperatures to a range of bombarding energies with the
help of the Gamow peak concept \cite{iliadis_2007_aa}, and then to
address this energy region with direct or indirect
measurements. Hoewever, there are pitfalls associated with using the
Gamow peak concept, as pointed out by
Refs.~\cite{newton_2007_aa,rauscher_2010_aa}. This procedure can only
be regarded as a rough estimate, partly because previously published
reaction rates have no rigorous statistical meaning and partly because
it is difficult to disentangle the impact of different nuclear physics
input parameters (resonance energies, strengths, partial widths,
spectroscopic factors, etc.) and their associated uncertainties on the
total reaction rate.

\begin{figure}
\begin{center}
\includegraphics[width=3.5in]{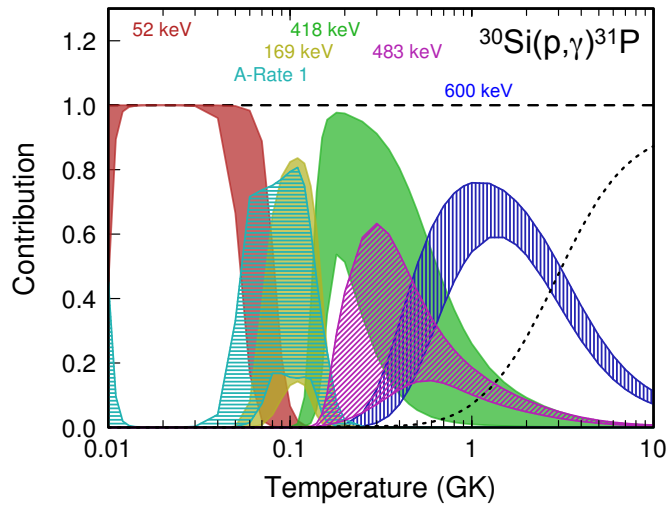}
\caption{Fractional contributions to the total $^{30}$Si(p,$\gamma$)$^{31}$P
reaction rate. Different colors indicate different contributions. The
vertical width of a band, indicating the uncertainty of a fractional
contribution, has a precise statistical meaning - a coverage
probability of 68\%. The results are obtained from the Monte Carlo
based method described in the text. Numbers at the top denote
center-of-mass energies of given resonances; ``A-Rate 1'' refers to
the non-resonant (direct capture) rate contribution; the dotted line
shows contributions of resonances with energies larger than 600 keV. From A.~E. Champagne, C. Iliadis, and R. Longland, AIP Advances 4, 041006 (2014).}
\label{fig:si30pg}
\end{center}
\end{figure}

The availability of Monte Carlo based reaction rates opens a new
window of opportunity. Since each nuclear physics input parameter is
sampled according to a physically motivated probability density
function \cite{longland_2010_aa}, the Monte Carlo sampling provides a
statistically rigorous coverage probability for each contribution. As
an example, consider Fig.~\ref{fig:si30pg}, showing the main
fractional contributions of individual observed or unobserved
resonances to the total $^{30}$Si(p,$\gamma$)$^{31}$P Monte Carlo
reaction rate. This reaction is of particular interest for
interpreting observed silicon isotopic ratios in nova candidate
presolar grains \cite{jose_2007_ab}. Different colors correspond to
contributions of different resonances, while the vertical width of
each band signifies a coverage probability of 68\%. Inspection of the
figure clearly identifies what needs to be measured in order to
improve the total reaction rate estimate, without resorting to the
Gamow peak concept. For example, at typical classical nova peak
temperatures near 300 MK the total rate is dominated by the
uncertainties in the contributions of the 418 keV and the 483 keV
resonances. The former has not been directly observed yet, while the
latter has been observed, albeit with a large uncertainty in the
experimental resonance strength.

Fractional reaction rate contributions, computed using the Monte Carlo
method, will play an important role for the design of future
measurements at existing or planned nuclear physics laboratories since
they identify the rate contributions to be measured and indicate the
degree of experimental precision required. 

\subsection{STARLIB: a New Nuclear Rate Library for Stellar Modeling}
\label{sec:starlib}

The solid black line in the top panel of Fig.~\ref{fig:ne22an}
represents a lognormal function that closely describes the
actual Monte Carlo reaction rate probability density shown as the red
histogram. It was found in Ref.~\cite{iliadis_2010_aa} that lognormal
distributions provide a useful approximation for the majority of the
reaction rate probability densities. This aspect is interesting
because, as discussed above, a lognormal function is defined by only two
parameters, $\mu$ and $\sigma$. The first parameter is related to the
{\it median rate} via $x_{med}=e^{\mu}$, while the second parameter is
related to the {\it factor uncertainty with respect to the median} via
$f.u.=e^{\sigma}$ (for a coverage probability of 68\%). Therefore, by
tabulating values for temperature, $T$, recommended rate, $x_{med}$,
and factor uncertainty, $f.u.$, the rate probability density function
can be computed conveniently by using Eq.~(\ref{lognormalpdf}) at any
temperature bounded by the table entries.

These ideas are crucial for the design of a next-generation reaction
rate library, called STARLIB \cite{sallaska_2013_aa}. Existing
libraries contain values of only two quantities, i.e., temperature and
recommended rate, either as analytical fit formulas (e.g., JINA
REACLIB or BDAT \cite{Cyburt:2010ey}) or in tabular format (e.g.,
BRUSLIB \cite{Aikawa:2005cn}). STARLIB also uses a tabular format,
providing temperature and recommended rate, but in addition lists the
factor uncertainty as a third parameter, which can be used for two
purposes. First, it provides an estimate for the rate uncertainty
since the coverage probability for rate values between
$x_{low}=e^\mu/e^\sigma=x_{med}/(f.u.)$ and $x_{high}=e^\mu
e^\sigma=x_{med}(f.u.)$ is 68\%. Second, the listed values for
$x_{med}$ and $f.u.$ determine the entire rate probability density
function, as discussed above. This aspect is important because it
allows for a convenient implementation of Monte Carlo based reaction
rates in nucleosynthesis studies, as will be discussed in the next
section. Experimental Monte Carlo based thermonuclear reaction rates
are so far available for about 70 nuclear reactions involving target
nuclei in the $A=14-40$ range. Experimental $\beta$-decay rates,
including their lognormal decay constant probability densities, are
easily incorporated into the structure of STARLIB
\cite{sallaska_2013_aa}.

A general-purpose nuclear reaction and decay library must also
encompass tens of thousands of nuclear interactions for which no
experimental information exists. For these reactions, STARLIB contains
theoretical rates that are computed using the nuclear reaction code
TALYS\footnote{See \texttt{http://www.talys.eu.}}. Reliable
uncertainties for theoretical reaction rates are difficult to assess
at present. Various claims have been made in the literature (``on
average within a factor of two''), which may be
optimistic. Traditionally, such uncertainties have been systematically
evaluated using reaction codes by exploring different sets of nuclear input models for each target and each reaction
channel. A
similar approach could be followed to estimate the uncertainties
affecting the TALYS rates. However, the present version of STARLIB
adopts a recommended factor of 10 uncertainty for any reaction rate
for which no experimental cross section information exists. This value
represents currently our best estimate based on experience. The factor
uncertainty, together with the recommended rate, is used to
compute the lognormal rate probability density for TALYS-based rates
in the same manner as for the experimental Monte Carlo rates discussed
above. One of the future goals is to replace many of the
theoretical rates with experimental Monte Carlo-based estimates. A
detailed discussion of the publicly available STARLIB reaction rate 
library\footnote{See \texttt{http://starlib.physics.unc.edu.}}
is given in Ref.~\cite{sallaska_2013_aa}.

\section{Monte Carlo Nucleosynthesis}
\label{sec:MCNucleo}
STARLIB contains the rate probability densities of all reactions in
the network. Therefore, the obvious next step is to employ these rates
in Monte Carlo nucleosynthesis studies. Notice that we are referring
to two different Monte Carlo procedures: the first is used to derive
reaction rates by {\it randomly sampling over the experimental nuclear
  physics input}, as described in the previous section (step 1), while
the second refers to estimating abundances by {\it randomly sampling
  over the reaction rates}, as will be discussed below (step 2). This
method is flexible, in the sense that all reaction rates can be varied
simultaneously, or specific groups of reactions can be studied
separately. Here we will concentrate on the former case. The general
strategy is to randomly sample the rates for every reaction in the network
and to compute a single nucleosynthesis model.
The procedure is repeated many times to collect an ensemble of
abundances from different reaction network runs, which can then be
further analyzed. It should be emphasized that pairs of corresponding
forward and reverse rates should {\it not} be sampled independently
since they are correlated according to the reciprocity theorem. The
proper sampling procedure in this case has been discussed in detail by
Ref.~\cite{sallaska_2013_aa}.

Monte Carlo nucleosynthesis studies have been performed previously
(e.g., \cite{Stoesz:2003vw,Parikhetal:2008uu,Roberts:2006va}), but did
not use statistically meaningful rate probability density functions
derived from experimental nuclear physics input. Instead, previous studies, as is
common practice, assigned arbitrary ``enhancement'' factors to the
rates. In most cases, these enhancement factors were globally defined
by identifying the type of reaction rate (e.g., whether from
experimental data, or purely from theory). In particular, these
enhancement factors were assumed to be independent of temperature.
Given the discussion of Monte Carlo reaction rates in
Sec.~\ref{sec:MCRates}, it is clear that this assumption is not valid
in general. Rather, the rate uncertainties display a strong
temperature-dependence arising from different resonance contributions (see, for example, Fig.~\ref{fig:ca40ag})
and hence the temperature dependence must be considered carefully in
the sampling procedure.

One choice for performing Monte Carlo nucleosynthesis studies (step 2)
is to utilize directly the random samples of the Monte Carlo reaction rates (step 1).
The advantage of this choice is that the individual rate samples are directly based on the
nuclear physics input and thus will account for all possible
behaviors of reaction rates as a function of
temperature. The disadvantage is that it requires a considerable amount
of effort, along with a detailed knowledge of all nuclear physics input
not widely available to users. 
Therefore, we describe below a reaction rate sampling method that is simpler
to implement and agrees well with more complex choices \cite{Longland:2012kv}.

\subsection{Reaction Rate Sampling}
\label{sec:sampling}
Most reactions for which we have computed Monte Carlo-based rates so
far are dominated by resonances. In the absence of interference
effects, the individual resonant contributions are summed incoherently
to obtain the total reaction rate. Therefore, reactions that involve a
large number of resonances with uncorrelated uncertainties may exhibit
complex rate variations from random sample to sample. One important
physical constraint is that these reaction rates must be a smooth
function of temperature owing to the convolution of the nuclear
reaction cross section with the thermal energy distribution of the
particles in the stellar plasma.

The two parameters, lognormal $\mu$ and $\sigma$, define the
approximate reaction rate probability density function. These
parameters form the basis of our sampling scheme. For a lognormal
probability density, a random rate sample, $x(T)$, at a specific
temperature, $T$, is computed from
\begin{equation}
  \label{eq:rate}
  x(T) = e^{\mu(T)} \cdot e^{p(T) \sigma(T)} = x(T)_{med} \cdot  (f.u.(T))^{p(T)}
\end{equation}
where $x_{med}(T)$ = $e^{\mu(T)}$ and $f.u.(T)$ = $e^{\sigma(T)}$ are the median rate and factor uncertainty (Sec.~\ref{sec:starlib}), respectively, and
$p(T)$ is a random variable that is normally distributed, i.e.,
according to a Gaussian distribution with an expectation value of zero
and standard deviation of unity. Note that $(f.u.(T))^{p(T)}$, and not
$p(T)$, is the factor by which the sampled reaction rate is modified
from its median value. The former factor depends on temperature
through both $(f.u.(T))$ and $p(T)$. With the lognormal
parameters, $\mu(T)$ and $\sigma(T)$, given in STARLIB, 
sampling over reaction rates becomes a simple task of assuming an
appropriate sampling scheme for the random variable $p(T)$.

\begin{figure}
\centering
\includegraphics[width=0.5\textwidth]{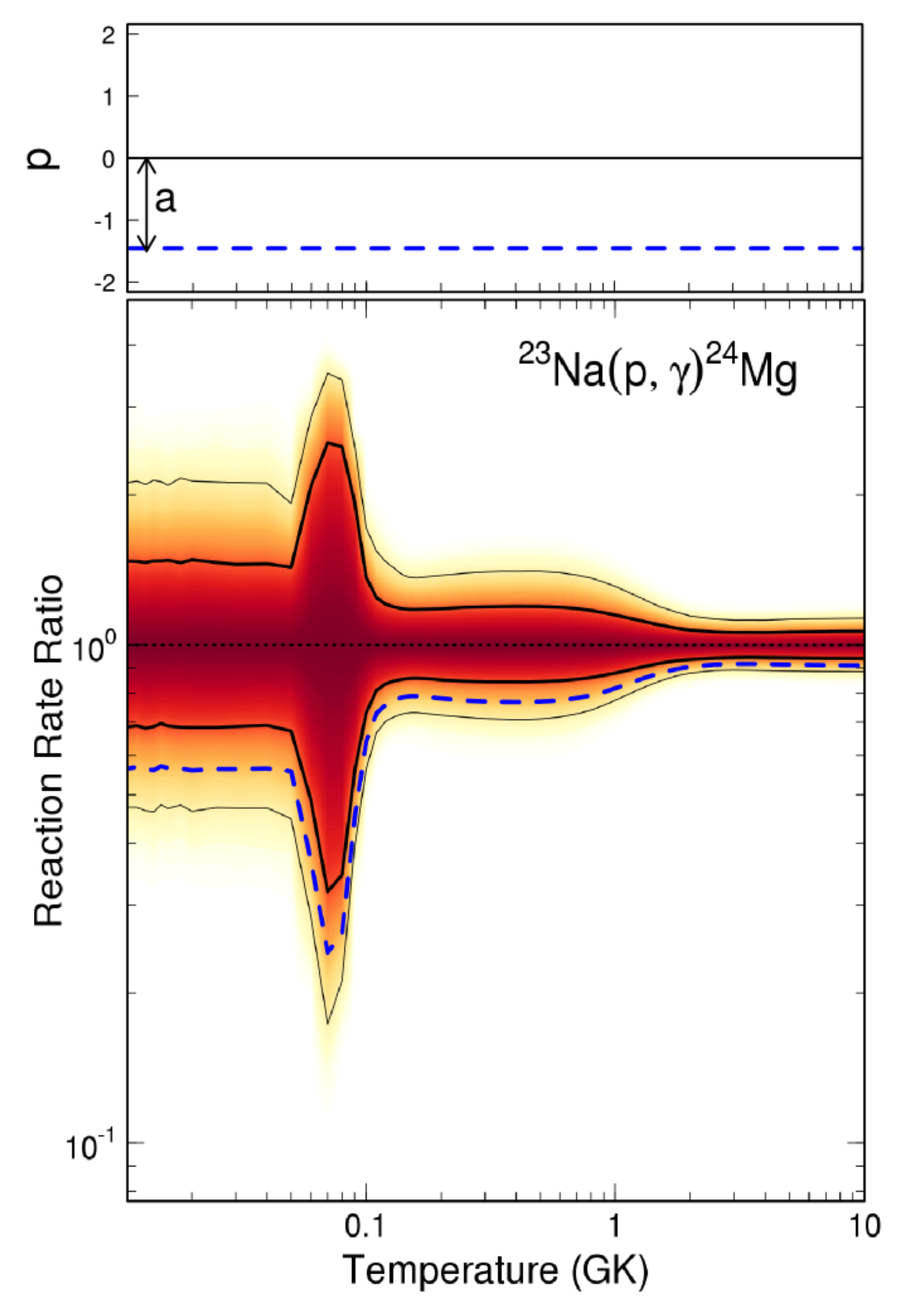}
\caption{
Reaction rate ratio (i.e., normalized to the recommended rate) for the
$^{23}$Na(p,$\gamma$)$^{24}$Mg reaction, obtained using the Monte
Carlo method discussed in Sec.~\ref{sec:newmethod}. The color
indicates the coverage probability of the rate density function. The
thick and thin black lines correspond to 68\% and 95\% rate
uncertainties, respectively, similar to Fig.~\ref{fig:ca40ag}. The
dashed blue line represents a single reaction rate sample obtained
with Eq.~(\ref{eq:rate}) and $p(T)$=$a$. From A.~E. Champagne, C. Iliadis, and R. Longland, AIP Advances 4, 041006 (2014).}
\label{fig:Sample}
\end{figure}

The simplest parameterization for $p(T)$ is obtained by assuming that
it is independent of temperature, i.e., $p(T)$=$a$, where $a$ is
sampled from a normal distribution. This parameterization was found to
reproduce the abundance uncertainties arising from more complex
sampling schemes \cite{Longland:2012kv}. Even in this simplest case
the rate uncertainty factor, given by $e^{a\, \sigma(T)}$ in
Eq.~(\ref{eq:rate}), is still temperature dependent. This temperature
dependence was disregarded in previous Monte Carlo studies (e.g.,
Ref.~\cite{Parikhetal:2008uu}) that employed a constant,
temperature-independent value for the uncertainty
factor. Figure~\ref{fig:Sample} illustrates this point, showing that a
uniform value of $p(T)$=$a$ produces a rate sample with a
temperature-dependent uncertainty. In the following we will refer to
$p(T)$=$a$ as the {\it rate variation factor}.

\subsection{Nucleosynthesis Studies}
\label{sec:MCNucleoSynthesis}
A Monte Carlo study of the nucleosynthesis can be performed following these steps: (i) for each (forward) reaction, the normally
distributed variables, $p_i$, are sampled independently; (ii) the
rates obtained are used to compute the nucleosynthesis for a single
post-processing network run; (iii) steps (i) and (ii) are repeated
a sufficient number of times to obtain an ensemble of final nucleosynthesis abundance
yields. The Monte Carlo procedure has major advantages compared to
varying rates one-by-one in sequential network runs. On the one hand,
it is straightforward to derive from the ensemble of final Monte Carlo
abundances the recommended values and associated uncertainties. One
choice is to adopt the 16th, 50th, and 84th percentiles as the low,
recommended, and high abundance, respectively, similar to the recipe
discussed in connection with reaction rates
(Sec.~\ref{sec:newmethod}). On the other hand, the impact of a given
reaction rate uncertainty on the nucleosynthesis can be easily
quantified by storing the values of $p_i$ for each sample reaction
network run. A scatter plot of the final abundance for a given nuclide
versus the sampled value of $p_i$ can then be investigated for
correlations.

In order to illustrate these points, we will apply in the following
the Monte Carlo method to big bang nucleosynthesis. Observations of
primordial $^4$He, $^2$H and $^7$Li abundances have reached an
unprecedented level of precision. Therefore, reliable predicted
abundances are needed before the observations can be confronted with
theory. Such studies have interesting implications for testing
standard or non-standard big bang models, since big bang
nucleosynthesis represents the earliest milestone of known physics
when we look back in time. The nucleosynthesis problem is also
well-defined, in the sense that the number of nuclear reactions in the
network is relatively small. Furthermore, the temperature and density
evolution during the big bang can be calculated from first principles
and hence the results are independent of convection, mass loss,
opacities, magnetic fields, etc., which introduce additional 
uncertainties to stellar nucleosynthesis predictions. Numerous studies
(see, e.g., Ref.~\cite{cyburt_2008_aa}) have shown
that the predicted big bang abundances of the light elements, for
example, $^2$H and $^4$He, agree with the observations. The sole
exception is $^7$Li, which is overproduced by a factor of three
relative to observations. The $^7$Li problem represents the central
unsolved puzzle of the nucleosynthesis in the early universe. A
solution is actively sought either in observation, nuclear physics, or
new physics beyond the standard model \cite{fields_2011_aa}.

\begin{figure}
\centering
\includegraphics[width=1.0\textwidth]{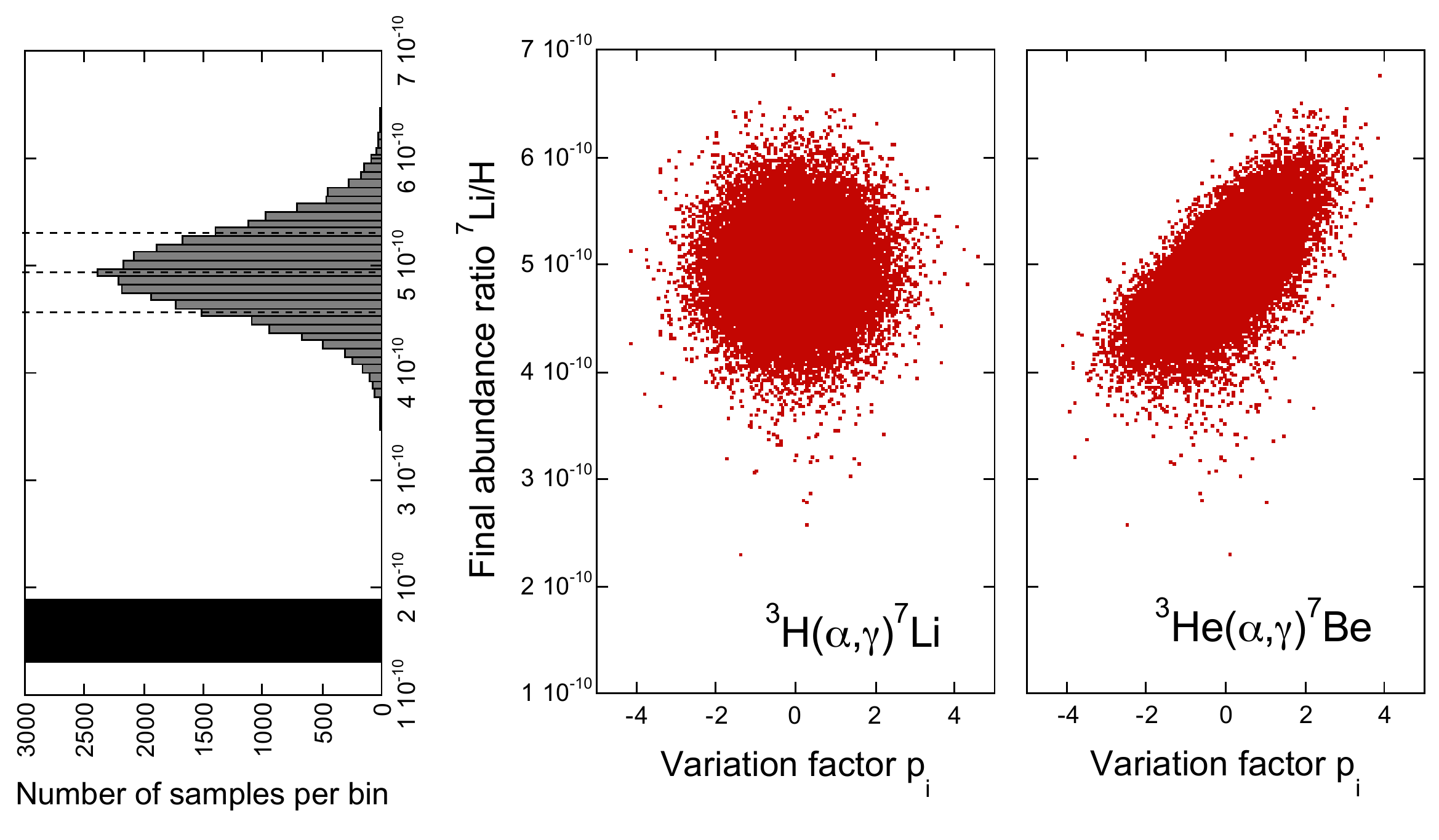}
\caption{
Results of a Monte Carlo study of big bang nucleosynthesis, using
30,000 reaction network runs. (Left) Histogram of final $^7$Li/H
number abundance ratio; the dashed lines indicate the 16th, 50th, and
84th percentiles of the abundance distribution; the observed range is
indicated by the black bar. (Middle) Final $^7$Li/H number abundance
ratio versus variation factor of the $^3$H($\alpha$,$\gamma$)$^7$Li
rate. (Right) Final $^7$Li/H number abundance ratio versus variation
factor of the $^3$He($\alpha$,$\gamma$)$^7$Be rate. The scatter plot
in the middle panel shows no correlation, while a strong correlation
is apparent in the right panel. The projection of either scatter plot
on the ordinate results in the panel shown on the left.}
\label{fig:BBMC}
\end{figure}

Results from the Monte Carlo procedure are shown in
Fig.~\ref{fig:BBMC}. The reaction network includes 59 nuclides between
the neutron and Na, and consists of 424 nuclear interactions
\cite{Coc:2011jh}. It is much larger than in other studies of standard
big bang nucleosynthesis because it is designed to study weak flows
into the CNO region. For this network, the rates of all reactions,
except for reverse reactions (Sec.~\ref{sec:MCNucleo}), are
independently sampled. In other words, each rate is multiplied by a
temperature-independent random variation factor, $p_i$, and thus is
sampled according to a temperature-dependent lognormal probability
density function (Eq.~\ref{eq:rate}). The sampling is repeated for
30,000 reaction network runs. The histogram of all 30,000 final number
abundance ratios $^7$Li/H is displayed in the left panel of
Fig.~\ref{fig:BBMC}. Note that $^7$Li is mainly produced in the big
bang as $^7$Be, which quickly decays to $^7$Li. This indirect big bang
synthesis of $^7$Li exceeds the direct synthesis by a factor of about
20. The dashed lines represent the 16th, 50th, and 84th percentiles,
which amount to $^7$Li/H = $4.56$ $\times$ $10^{-10}$, $4.94$ $\times$
$10^{-10}$, and $5.34$ $\times$ $10^{-10}$, respectively. These values
are based on our best estimates for the probability density functions
of all reaction rates in the network, including all {\it known statistical
  and systematic effects} (Sec.~\ref{sec:newmethod}). As already
noted, the observations result in a factor of three smaller $^7$Li/H
ratio, indicated by the black bar in the lower part of the left
panel. Thus it is unlikely that the solution of the $^7$Li problem
will be found in uncertain nuclear physics.  This conjecture
disregards {\it unknown systematic effects}.

The next question at hand is: which reaction rates have the strongest
impact on the predicted $^7$Li/H ratio?  Once identified, these
reactions can be subjected to further scrutiny and the experimental
data can be inspected for previously unaccounted systematic
effects. The Monte Carlo results presented above contain an answer to
this question since the variation factors, $p_i$, of each reaction are
recorded for each of the 30,000 network calculations. Consider again
Fig.~\ref{fig:BBMC}, displaying the final $^7$Li/H ratios of all
30,000 network runs versus the sampled rate variation factors, $p_i$,
of $^3$H($\alpha$,$\gamma$)$^7$Li (middle panel) and
$^3$He($\alpha$,$\gamma$)$^7$Be (right panel). The projection of any
of these scatter plots onto the ordinate results in the (same)
histogram shown on the left, which was already discussed above. Again,
the spread along the y-direction is caused by the combined
uncertainties of all reaction rates in the network. The impact of a
given reaction rate on the final abundance of a given nuclide, $X_f$,
is apparent in the scatter plot: if the variation in $p_i$ results in
a flat distribution of $X_f$, we can conclude that the given rate and
the given abundance are uncorrelated. This is the case depicted in the
middle panel. On the other hand, if the variation in $p_i$ results in
a systematic change of $X_f$, the given rate and the given abundance
are correlated. This is the situation shown in the right panel.

The $^3$H($\alpha$,$\gamma$)$^7$Li and $^3$He($\alpha$,$\gamma$)$^7$Be
reaction rates have currently uncertainty factors of $f.u.$ = $1.07$
(i.e., 7\%) and $f.u.$ = $1.05$ (i.e., 5\%), respectively, at big bang
nucleosynthesis temperatures. The strong impact of the latter reaction
is easily explained because it directly produces $^7$Be. The former
reaction bypasses the synthesis of $^7$Be and, furthermore, does not
contribute to the direct synthesis of $^7$Li because the produced
$^7$Li nuclei are quickly destroyed by the strong subsequent
$^7$Li(p,$\alpha$)$\alpha$ reaction.

The identification of $^3$He($\alpha$,$\gamma$)$^7$Be as the most
important reaction impacting the nucleosynthesis of $^7$Li is not
surprising and could have been guessed without Monte Carlo procedures
because the standard big bang nucleosynthesis network includes
nuclides up to $^7$Be and hence contains few reactions only. However,
reaction networks for other astrophysical scenarios, for example,
advanced and explosive burning stages in massive stars, binary star
explosions (type Ia supernovae, classical novae, type I x-ray bursts),
s-process, r-process, p-process, etc., contain several thousand
nuclear interactions. Therefore, the identification of the most
important reactions that impact a given isotopic abundance is not
as obvious at first sight.

We have found the following procedure useful in several
applications. Recall that the Monte Carlo study outputs the final
abundances and rate variation factors for each sample network run. For
each combination of nuclide and reaction a correlation factor is
computed based on the scatter plot of final abundance versus rate
variation factor. For a given nuclide, all reactions are then sorted
in descending order according to the magnitude of the correlation
factor. The reactions at the top of the list are the ones whose
current rate uncertainties have the strongest impact on the nuclidic
abundance.

\begin{figure}
\centering
\includegraphics[width=1.0\textwidth]{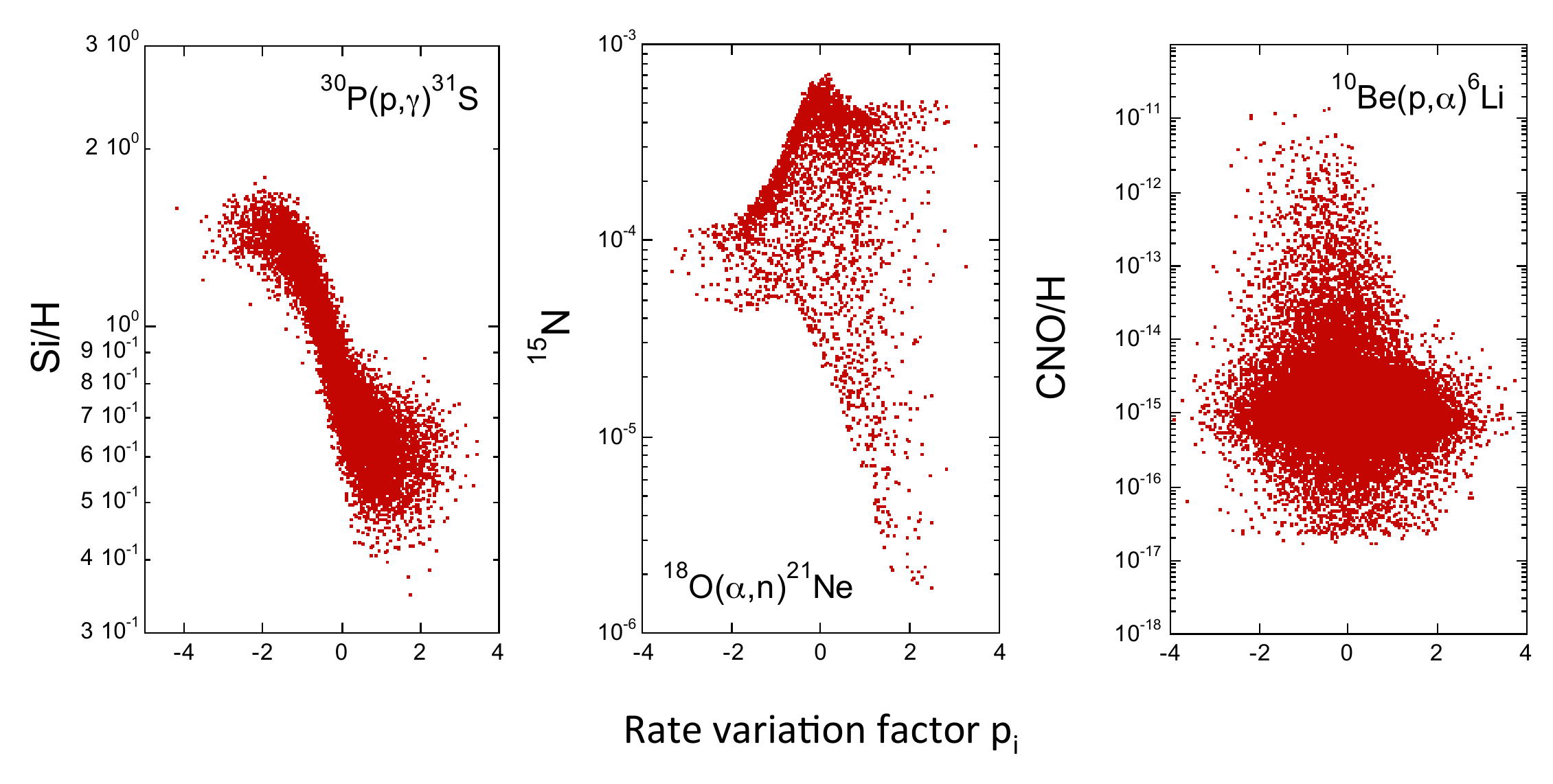}
\caption{Results of Monte Carlo nucleosynthesis studies. (Left) Final mass
fraction ratio of Si/H versus rate variation factor for
$^{30}$P(p,$\gamma$)$^{31}$S at the end of a classical nova simulation
involving a 1.35~M$_\odot$ white dwarf. (Middle) Final mass fraction
of $^{15}$N versus rate variation factor for
$^{18}$O($\alpha$,n)$^{21}$Ne from from supernova shock-induced nucleosynthesis. 
(Right) Final number abundance ratio CNO/H versus rate
variation factor for $^{10}$Be(p,$\alpha$)$^{6}$Li in the context of
an extended big bang network (391 reactions).}
\label{fig:BBMC2}
\end{figure}

When a linear correlation is known to be significant, the most
commonly used correlation factor is Pearson's $r$
\cite{press_1992_aa}, with values ranging from +1 to $-$1. It is
defined for two variables, $x_k$ and $y_k$, by the expression
\begin{equation}
  {
   r = \frac{\sum_k (x_k - \bar{x})(y_k - \bar{y})}{\sqrt{\sum_k (x_k - \bar{x})^2} \sqrt{\sum_k (y_k - \bar{y})^2}}
   }
\label{eq:pearson}
\end{equation}
where $\bar{x}$ and $\bar{y}$ denote the mean values. If all the data
lie on a perfect straight line, and the line has a positive slope, $r$
= $+1$; for a negative slope, $r$ = $-1$. These values hold
independently of the magnitude of the slope. A value of $r$ near zero
indicates that no correlation exists. There are a number of
disadvantages when using Pearson's $r$ in connection with scatter
plots of final abundance versus rate variation factor. First, it is
well known that $r$ is a poor statistic for deciding {\it whether} an
observed correlation is statistically significant. Second, Pearson's
$r$ is a measure for a {\it linear} correlation between two
variables. However, actual scatter plots frequently reveal non-linear
correlations. This is demonstrated in Fig.~\ref{fig:BBMC2}, displaying
a number of examples we have obtained for various scenarios: mass
fraction ratio Si/H versus rate variation factor for
$^{30}$P(p,$\gamma$)$^{31}$S at the end of a classical nova simulation
involving a 1.35~M$_\odot$ white dwarf \cite{Kelly:2013jj} (left panel); final $^{15}$N
mass fraction versus rate variation factor for
$^{18}$O($\alpha$,n)$^{21}$Ne from supernova shock-induced nucleosynthesis. This occurs in the innermost region of the helium shell in a 15 M$_{\odot}$ massive star explosion model \cite{MEY13}
(middle panel); and number abundance ratio CNO/H versus rate variation
factor for $^{10}$Be(p,$\alpha$)$^{6}$Li in the context of an extended
big bang nucleosynthesis network (right panel). None of these examples exhibit a
linear correlation.

A better choice is the Spearman rank-order correlation coefficient
($r_s$), which quantifies how well the relationship between two
variables is described by a monotonic function. It is obtained from
Eq.~\ref{eq:pearson} when the actual values of $x_i$ and $y_i$ are
replaced by their ranks. Values of $r_s$ = $+1$ or $r_s$ = $-1$ are
obtained if the relationship between the two variables is given by a
perfectly increasing or decreasing monotonic function,
respectively. The Spearman coefficient is also more robust compared to
Pearson's $r$ in connection with outliers, in a similar sense that the
median value of a distribution is more robust to outliers than the
mean value.

For the scatter plots shown in Fig.~\ref{fig:BBMC}, the Spearman
coefficients amount to $r_s$ = $0.012$ (middle panel) and $+0.69$
(right panel), verifying that no correlation exists between the
$^7$Li/H ratio and the $^3$H($\alpha$,$\gamma$)$^7$Li rate, and that
the $^7$Li/H ratio is correlated with the
$^3$He($\alpha$,$\gamma$)$^7$Be rate. 
For the scatter plot shown in the left panel of
Fig.~\ref{fig:BBMC2}, we obtain $r_s$ = $-0.86$. The Spearman
coefficient correctly predicts a strong correlation because the
relationship between the displayed variables is monotonic. For the
middle panel, we obtain a small value of $r_s$ = $0.05$ and hence this
case could be easily overlooked if the impact of reaction rates is
quantified and ranked according to the magnitude of the correlation
coefficient. Clearly, one has to be careful because the Spearman
coefficient is not designed to quantify such a non-uniform,
two-dimensional pattern.

Finally, for the right panel in Fig.~\ref{fig:BBMC2} we obtain a value
of $r_s$ = $-0.213$, indicating a significant negative correlation
between the number abundance ratio CNO/H and the
$^{10}$Be(p,$\alpha)^6$Li reaction rate for an extended big bang
nucleosynthesis network. The Monte Carlo method results in a
primordial range of CNO/H = $(4.94 - 28.5)$ $\times$ $10^{-16}$ (68\%
coverage probability). It confirms earlier results \cite{Iocco:2007jv}, but provides in addition a meaningful uncertainty. Notice the tail towards higher values, CNO/H
$>$ $10^{-13}$, for which a small, but finite, probability is obtained
(2\%). This result could imply that some of the first-generation massive stars \cite{ekstrom_2008_aa} do not need to self-produce as much CNO nuclei to meet their nuclear energy generation requirements. 
This region of the scatter plot is caused by the combination of
higher sampled rates for reactions that connect to the CNO region,
chiefly $^8$Li(t,n)$^{10}$Be and $^{10}$Be($\alpha$,n)$^{13}$C, and
lower sampled rates for reactions that return matter to the lightest
nuclides, mainly $^{10}$Be(p,$\alpha$)$^7$Li and
$^{10}$Be(p,t)2$\alpha$. Since these four reactions involving
radioactive $^{10}$Be have not been measured in the laboratory yet,
the rates were estimated from theory (using TALYS). Thus we assign a
factor uncertainty of $f.u.$ = $100$ to these reaction rates, which is
a reasonable estimate for reactions among light nuclei at low
energies. The identification of this potentially new path from Li
towards the CNO region via $^{10}$Be could not have been made if the
rates were varied individually one-by-one in sequential network runs
(i.e., without using the Monte Carlo method): individual variations of
these rates by a factor of 1,000 yield at most a 30\% increase in CNO
production!

\section{Summary and Conclusions}

Statistical methods are necessary to improve estimates of
both thermonuclear reaction rates and nucleosynthesis predicted by
nuclear reaction networks. Experimental reaction rates can be
estimated by using a Monte Carlo method once appropriate probability
density functions are adopted for each nuclear physics input
quantity. For example, resonance energies are best described by a
Gaussian probability density, while for resonance strengths a
lognormal probability density is more appropriate. Unobserved
resonances of unknown strength can be easily incorporated into this
framework by assuming a Porter-Thomas probability density. The random
sampling over all nuclear input parameters produces in most cases a
lognormal (output) reaction rate probability density. This function
provides statistically rigorous recommended (median) reaction rates
and factor uncertainties. We also discussed the usefulness of the
Monte Carlo method for estimating the fractional contributions to the
total reaction rate. Such calculations are important for the design of
experiments at stable beam and radioactive ion beam facilities.

These results led directly to the construction of a next-generation
nuclear reaction library, STARLIB, containing reaction rates,
uncertainties, and rate probability densities for easy implementation
into reaction networks for stellar models. STARLIB contains the
necessary nuclear physics information to perform Monte Carlo
nucleosynthesis simulations. All reaction rates can be sampled
simultaneously, except for reverse reactions since these are related
to the corresponding forward reaction rates via the reciprocity
theorem. For the sampling of reaction rates, which are described by
lognormal probability densities, we introduce a Gaussian random
variable, $p_i$, for each each reaction, $i$. In the simplest case, it
can be assumed that this {\it rate variation factor} is independent of
temperature. The random factor by which a given sampled rate is
modified from its median value is given by $(f.u.)^{p_i}$, which is
temperature-dependent in any case through the factor uncertainty,
$f.u.$

Finally, we discussed how to assess the impact of reaction rate uncertainties on
the abundance of a given nuclide. Scatter plots displaying the final
abundance, $X_f$, versus the rate variation factor, $p_i$, of reaction
$i$ are useful for quantifying correlations. We provide several
examples that impact s-process neutron sources, core-collapse supernovae,
classical novae, and big bang nucleosynthesis. The challenge is to
identify which reactions, among typically several thousand in an
extended network, have the largest impact on a given
abundance. Useful results are obtained with the Spearman rank-order
correlation coefficient, which is designed to quantify correlations
for a monotonic relationship between two variables. We also gave
examples for more complicated correlations that are not easily
identified on the basis of the Spearman coefficient alone.

\section{Acknowlegements}
This work was supported in part by the National Science Foundation under award number AST-1008355 and by the U.S. Department of Energy under Contracts No. DE-FG02-97ER41041 and DE-FG02-97ER41042.

\section*{References}

\bibliographystyle{unsrt}

\bibliography{master}

\begin{thebibliography}{10}

\bibitem{abazajian_2009_aa}
K.~N. {Abazajian}, J.~K. {Adelman-McCarthy}, M.~A. {Ag{\"u}eros}, S.~S.
  {Allam}, C.~{Allende Prieto}, D.~{An}, K.~S.~J. {Anderson}, S.~F. {Anderson},
  J.~{Annis}, N.~A. {Bahcall}, and et~al.
\newblock {The Seventh Data Release of the Sloan Digital Sky Survey}.
\newblock {\em \apjs}, 182:543--558, June 2009.

\bibitem{rau_2009_aa}
A.~{Rau}, S.~R. {Kulkarni}, N.~M. {Law}, J.~S. {Bloom}, D.~{Ciardi}, G.~S.
  {Djorgovski}, D.~B. {Fox}, A.~{Gal-Yam}, C.~C. {Grillmair}, M.~M. {Kasliwal},
  P.~E. {Nugent}, E.~O. {Ofek}, R.~M. {Quimby}, W.~T. {Reach}, M.~{Shara},
  L.~{Bildsten}, S.~B. {Cenko}, A.~J. {Drake}, A.~V. {Filippenko}, D.~J.
  {Helfand}, G.~{Helou}, D.~A. {Howell}, D.~{Poznanski}, and M.~{Sullivan}.
\newblock {Exploring the Optical Transient Sky with the Palomar Transient
  Factory}.
\newblock {\em \pasp}, 121:1334--1351, December 2009.

\bibitem{law_2009_aa}
N.~M. {Law}, S.~R. {Kulkarni}, R.~G. {Dekany}, E.~O. {Ofek}, R.~M. {Quimby},
  P.~E. {Nugent}, J.~{Surace}, C.~C. {Grillmair}, J.~S. {Bloom}, M.~M.
  {Kasliwal}, L.~{Bildsten}, T.~{Brown}, S.~B. {Cenko}, D.~{Ciardi},
  E.~{Croner}, S.~G. {Djorgovski}, J.~{van Eyken}, A.~V. {Filippenko}, D.~B.
  {Fox}, A.~{Gal-Yam}, D.~{Hale}, N.~{Hamam}, G.~{Helou}, J.~{Henning}, D.~A.
  {Howell}, J.~{Jacobsen}, R.~{Laher}, S.~{Mattingly}, D.~{McKenna},
  A.~{Pickles}, D.~{Poznanski}, G.~{Rahmer}, A.~{Rau}, W.~{Rosing}, M.~{Shara},
  R.~{Smith}, D.~{Starr}, M.~{Sullivan}, V.~{Velur}, R.~{Walters}, and
  J.~{Zolkower}.
\newblock {The Palomar Transient Factory: System Overview, Performance, and
  First Results}.
\newblock {\em \pasp}, 121:1395--1408, December 2009.

\bibitem{koch_2010_aa}
D.~G. {Koch}, W.~J. {Borucki}, G.~{Basri}, N.~M. {Batalha}, T.~M. {Brown},
  D.~{Caldwell}, J.~{Christensen-Dalsgaard}, W.~D. {Cochran}, E.~{DeVore},
  E.~W. {Dunham}, T.~N. {Gautier}, III, J.~C. {Geary}, R.~L. {Gilliland},
  A.~{Gould}, J.~{Jenkins}, Y.~{Kondo}, D.~W. {Latham}, J.~J. {Lissauer},
  G.~{Marcy}, D.~{Monet}, D.~{Sasselov}, A.~{Boss}, D.~{Brownlee},
  J.~{Caldwell}, A.~K. {Dupree}, S.~B. {Howell}, H.~{Kjeldsen}, S.~{Meibom},
  D.~{Morrison}, T.~{Owen}, H.~{Reitsema}, J.~{Tarter}, S.~T. {Bryson}, J.~L.
  {Dotson}, P.~{Gazis}, M.~R. {Haas}, J.~{Kolodziejczak}, J.~F. {Rowe}, J.~E.
  {Van Cleve}, C.~{Allen}, H.~{Chandrasekaran}, B.~D. {Clarke}, J.~{Li}, E.~V.
  {Quintana}, P.~{Tenenbaum}, J.~D. {Twicken}, and H.~{Wu}.
\newblock {Kepler Mission Design, Realized Photometric Performance, and Early
  Science}.
\newblock {\em \apjl}, 713:L79--L86, April 2010.

\bibitem{batalha_2010_aa}
N.~M. {Batalha}, W.~J. {Borucki}, D.~G. {Koch}, S.~T. {Bryson}, M.~R. {Haas},
  T.~M. {Brown}, D.~A. {Caldwell}, J.~R. {Hall}, R.~L. {Gilliland}, D.~W.
  {Latham}, S.~{Meibom}, and D.~G. {Monet}.
\newblock {Selection, Prioritization, and Characteristics of Kepler Target
  Stars}.
\newblock {\em \apjl}, 713:L109--L114, April 2010.

\bibitem{paxton_2011_aa}
B.~{Paxton}, L.~{Bildsten}, A.~{Dotter}, F.~{Herwig}, P.~{Lesaffre}, and
  F.~{Timmes}.
\newblock {Modules for Experiments in Stellar Astrophysics (MESA)}.
\newblock {\em \apjs}, 192:3--+, January 2011.

\bibitem{paxton_2013_aa}
B.~{Paxton}, M.~{Cantiello}, P.~{Arras}, L.~{Bildsten}, E.~F. {Brown},
  A.~{Dotter}, C.~{Mankovich}, M.~H. {Montgomery}, D.~{Stello}, F.~X. {Timmes},
  and R.~{Townsend}.
\newblock {Modules for Experiments in Stellar Astrophysics (MESA): Planets,
  Oscillations, Rotation, and Massive Stars}.
\newblock {\em \apjs}, 208:4, September 2013.

\bibitem{balantekin_2014_aa}
A.~B. {Balantekin}, J.~{Carlson}, D.~J. {Dean}, G.~M. {Fuller}, R.~J.
  {Furnstahl}, M.~{Hjorth-Jensen}, R.~V.~F. {Janssens}, B.-A. {Li},
  W.~{Nazarewicz}, F.~M. {Nunes}, W.~E. {Ormand}, S.~{Reddy}, and B.~M.
  {Sherrill}.
\newblock {Nuclear theory and science of the facility for rare isotope beams}.
\newblock {\em Modern Physics Letters A}, 29:30010, April 2014.

\bibitem{cesaratto_2010_aa}
J.~M. {Cesaratto}, A.~E. {Champagne}, T.~B. {Clegg}, M.~Q. {Buckner}, R.~C.
  {Runkle}, and A.~{Stefan}.
\newblock {Nuclear astrophysics studies at LENA: The accelerators}.
\newblock {\em Nuclear Instruments and Methods in Physics Research A},
  623:888--894, November 2010.

\bibitem{longland_2006_aa}
R.~{Longland}, C.~{Iliadis}, A.~E. {Champagne}, C.~{Fox}, and J.~R. {Newton}.
\newblock {Nuclear astrophysics studies at the LENA facility: The
  {$\gamma$}-ray detection system}.
\newblock {\em Nuclear Instruments and Methods in Physics Research A},
  566:452--464, October 2006.

\bibitem{caughlan_1988_aa}
G.~R. {Caughlan} and W.~A. {Fowler}.
\newblock {Thermonuclear Reaction Rates V}.
\newblock {\em Atomic Data and Nuclear Data Tables}, 40:283, 1988.

\bibitem{angulo_1999_aa}
C.~Angulo, M.~Arnould, M.~Rayet, P.~Descouvemont, D.~Baye, C.~Leclercq-Willain,
  A.~Coc, S.~Barhoumi, P.~Aguer, C.~Rolfs, R.~Kunz, J.W. Hammer, A.~Mayer,
  T.~Paradellis, S.~Kossionides, C.~Chronidou, K.~Spyrou, S.~Degl'Innocenti,
  G.~Fiorentini, B.~Ricci, S.~Zavatarelli, C.~Providencia, H.~Wolters,
  J.~Soares, C.~Grama, J.~Rahighi, A.~Shotter, and M.~Lamehi Rachti.
\newblock A compilation of charged-particle induced thermonuclear reaction
  rates.
\newblock {\em Nuclear Physics A}, 656(1):3 -- 183, 1999.

\bibitem{iliadis_2001_aa}
C.~{Iliadis}, J.~M. {D'Auria}, S.~{Starrfield}, W.~J. {Thompson}, and
  M.~{Wiescher}.
\newblock {Proton-induced Thermonuclear Reaction Rates for A=20-40 Nuclei}.
\newblock {\em \apjs}, 134:151--171, May 2001.

\bibitem{Smith:2002uk}
D~L Smith, D~G Naberejnev, and L~A Van~Wormer.
\newblock {Large errors and severe conditions}.
\newblock {\em Nuclear Instruments and Methods A}, 488:342--361, July 2002.

\bibitem{longland_2010_aa}
R.~{Longland}, C.~{Iliadis}, A.~E. {Champagne}, J.~R. {Newton}, C.~{Ugalde},
  A.~{Coc}, and R.~{Fitzgerald}.
\newblock {Charged-particle thermonuclear reaction rates: I. Monte Carlo method
  and statistical distributions}.
\newblock {\em Nuclear Physics A}, 841:1--30, October 2010.

\bibitem{iliadis_2010_aa}
C.~{Iliadis}, R.~{Longland}, A.~E. {Champagne}, A.~{Coc}, and R.~{Fitzgerald}.
\newblock {Charged-particle thermonuclear reaction rates: II. Tables and graphs
  of reaction rates and probability density functions}.
\newblock {\em Nuclear Physics A}, 841:31--250, October 2010.

\bibitem{iliadis_2010_ab}
C.~{Iliadis}, R.~{Longland}, A.~E. {Champagne}, and A.~{Coc}.
\newblock {Charged-particle thermonuclear reaction rates: III. Nuclear physics
  input}.
\newblock {\em Nuclear Physics A}, 841:251--322, October 2010.

\bibitem{iliadis_2010_ac}
C.~{Iliadis}, R.~{Longland}, A.~E. {Champagne}, and A.~{Coc}.
\newblock {Charged-particle thermonuclear reaction rates: IV. Comparison to
  previous work}.
\newblock {\em Nuclear Physics A}, 841:323--388, October 2010.

\bibitem{Longland:2012ix}
R~Longland, C~Iliadis, and A~Karakas.
\newblock {Reaction rates for the s-process neutron source $^{22}$Ne +
  $\alpha$}.
\newblock {\em Physical Review C}, 85(6):065809, June 2012.

\bibitem{Iliadis:1997dh}
C~Iliadis.
\newblock {Proton single-particle reduced widths for unbound states}.
\newblock {\em Nuclear Physics A}, 618(1-2):166--175, May 1997.

\bibitem{porter_1956_aa}
C~E Porter and R.G. Thomas.
\newblock {Fluctiations of nuclear reaction widths}.
\newblock {\em Phys. Rev.}, page 483, March 1956.

\bibitem{pogrebnyak_2013_aa}
I~Pogrebnyak, C~Howard, C~Iliadis, R~Longland, and G~E Mitchell.
\newblock {Mean proton and $\alpha$-particle reduced widths of the
  Porter-Thomas distribution and astrophysical applications}.
\newblock {\em Physical Review C}, 88(1):015808, July 2013.

\bibitem{woosley_1973_aa}
S.~E. {Woosley}, W.~D. {Arnett}, and D.~D. {Clayton}.
\newblock {The Explosive Burning of Oxygen and Silicon}.
\newblock {\em \apjs}, 26:231--+, November 1973.

\bibitem{woosley_1992_aa}
S.~E. {Woosley} and R.~D. {Hoffman}.
\newblock {The alpha-process and the r-process}.
\newblock {\em \apj}, 395:202--239, August 1992.

\bibitem{timmes_1996_ab}
F.~X. {Timmes}, S.~E. {Woosley}, D.~H. {Hartmann}, and R.~D. {Hoffman}.
\newblock {The Production of 44Ti and 60Co in Supernovae}.
\newblock {\em \apj}, 464:332--+, June 1996.

\bibitem{the_1998_aa}
L.-S. {The}, D.~D. {Clayton}, L.~{Jin}, and B.~S. {Meyer}.
\newblock {Nuclear Reactions Governing the Nucleosynthesis of 44Ti}.
\newblock {\em \apj}, 504:500--+, September 1998.

\bibitem{hoffman_2010_aa}
R.~D. {Hoffman}, S.~A. {Sheets}, J.~T. {Burke}, N.~D. {Scielzo}, T.~{Rauscher},
  E.~B. {Norman}, S.~{Tumey}, T.~A. {Brown}, P.~G. {Grant}, A.~M. {Hurst},
  L.~{Phair}, M.~A. {Stoyer}, T.~{Wooddy}, J.~L. {Fisker}, and D.~{Bleuel}.
\newblock {Reaction Rate Sensitivity of $^{44}$Ti Production in Massive Stars
  and Implications of a Thick Target Yield Measurement of $^{40}$Ca({$\alpha$},
  {$\gamma$})$^{44}$Ti}.
\newblock {\em \apj}, 715:1383--1399, June 2010.

\bibitem{magkotsios_2010_aa}
G.~{Magkotsios}, F.~X. {Timmes}, A.~L. {Hungerford}, C.~L. {Fryer}, P.~A.
  {Young}, and M.~{Wiescher}.
\newblock {Trends in $^{44}$Ti and $^{56}$Ni from Core-collapse Supernovae}.
\newblock {\em \apjs}, 191:66--95, November 2010.

\bibitem{larsson_2011_aa}
J.~{Larsson}, C.~{Fransson}, G.~{{\"O}stlin}, P.~{Gr{\"o}ningsson},
  A.~{Jerkstrand}, C.~{Kozma}, J.~{Sollerman}, P.~{Challis}, R.~P. {Kirshner},
  R.~A. {Chevalier}, K.~{Heng}, R.~{McCray}, N.~B. {Suntzeff}, P.~{Bouchet},
  A.~{Crotts}, J.~{Danziger}, E.~{Dwek}, K.~{France}, P.~M. {Garnavich}, S.~S.
  {Lawrence}, B.~{Leibundgut}, P.~{Lundqvist}, N.~{Panagia}, C.~S.~J. {Pun},
  N.~{Smith}, G.~{Sonneborn}, L.~{Wang}, and J.~C. {Wheeler}.
\newblock {X-ray illumination of the ejecta of supernova 1987A}.
\newblock {\em \nat}, 474:484--486, June 2011.

\bibitem{grebenev_2012_aa}
S.~A. {Grebenev}, A.~A. {Lutovinov}, S.~S. {Tsygankov}, and C.~{Winkler}.
\newblock {Hard-X-ray emission lines from the decay of $^{44}$Ti in the remnant
  of supernova 1987A}.
\newblock {\em \nat}, 490:373--375, October 2012.

\bibitem{robertson_2012_aa}
D.~{Robertson}, J.~{G{\"o}rres}, P.~{Collon}, M.~{Wiescher}, and H.-W.
  {Becker}.
\newblock {New measurement of the astrophysically important
  $^{40}$Ca({$\alpha$},{$\gamma$})$^{44}$Ti reaction}.
\newblock {\em \prc}, 85(4):045810, April 2012.

\bibitem{iliadis_2007_aa}
C.~{Iliadis}.
\newblock {\em {Nuclear Physics of Stars}}.
\newblock Wiley-VCH Verlag, 2007.

\bibitem{newton_2007_aa}
J~Newton, C~Iliadis, A~Champagne, A~Coc, Y~Parpottas, and C~Ugalde.
\newblock {Gamow peak in thermonuclear reactions at high temperatures}.
\newblock {\em Physical Review C}, 75(4), April 2007.

\bibitem{rauscher_2010_aa}
T.~{Rauscher}.
\newblock {Relevant energy ranges for astrophysical reaction rates}.
\newblock {\em \prc}, 81(4):045807, April 2010.

\bibitem{jose_2007_ab}
Jordi Jose and Margarita Hernanz.
\newblock {The origin of presolar nova grains}.
\newblock {\em Meteorit. Planet. Sci.}, 42:1135--1143, May 2007.

\bibitem{sallaska_2013_aa}
A.~L. {Sallaska}, C.~{Iliadis}, A.~E. {Champange}, S.~{Goriely},
  S.~{Starrfield}, and F.~X. {Timmes}.
\newblock {STARLIB: A Next-generation Reaction-rate Library for Nuclear
  Astrophysics}.
\newblock {\em \apjs}, 207:18, July 2013.

\bibitem{Cyburt:2010ey}
Richard~H Cyburt, A~Matthew Amthor, Ryan Ferguson, Zach Meisel, Karl Smith,
  Scott Warren, Alexander Heger, R~D Hoffman, Thomas Rauscher, Alexander
  Sakharuk, Hendrik Schatz, F~K Thielemann, and Michael Wiescher.
\newblock {THE JINA REACLIB DATABASE: ITS RECENT UPDATES AND IMPACT ON TYPE-I
  X-RAY BURSTS}.
\newblock {\em The Astrophysical Journal Supplement Series}, 189(1):240--252,
  June 2010.

\bibitem{Aikawa:2005cn}
M~Aikawa, M~Arnould, S~Goriely, A~Jorissen, and K~Takahashi.
\newblock {BRUSLIB and NETGEN: the Brussels nuclear reaction rate library and
  nuclear network generator for astrophysics}.
\newblock {\em Astronomy {\&} Astrophysics}, 441(3):1195--1203, October 2005.

\bibitem{Stoesz:2003vw}
J~A Stoesz and F~Herwig.
\newblock {Oxygen isotopic ratios in first dredge-up red giant stars and
  nuclear reaction rate uncertainties revisited}.
\newblock {\em Mon. Not. R. Astr. Soc.}, 340:763, December 2003.

\bibitem{Parikhetal:2008uu}
A~Parikh~et al.
\newblock {The Effects of Variations in Nuclear Processes on Type I X-Ray Burst
  Nucleosynthesis}.
\newblock {\em Astrophys. J. Suppl.}, 2008.

\bibitem{Roberts:2006va}
L~F Roberts, R~W Hix, Smith MM, and Fisker~J LL.
\newblock {Monte Carlo simulations of Type I X-ray burst nucleosynthesis}.
\newblock In {\em Proceedings of Science}, pages 1--6, CERN, February 2006.

\bibitem{Longland:2012kv}
R~Longland.
\newblock {Recommendations for Monte Carlo nucleosynthesis sampling}.
\newblock {\em Astronomy {\&} Astrophysics}, 548:A30, November 2012.

\bibitem{cyburt_2008_aa}
R.~H. {Cyburt}, B.~D. {Fields}, and K.~A. {Olive}.
\newblock {An update on the big bang nucleosynthesis prediction for $^{7}$Li:
  the problem worsens}.
\newblock {\em Journal of Cosmology and Astro-Particle Physics}, 11:12--+,
  November 2008.

\bibitem{fields_2011_aa}
B.~D. {Fields}.
\newblock {The Primordial Lithium Problem}.
\newblock {\em Annual Review of Nuclear and Particle Science}, 61:47--68,
  November 2011.

\bibitem{Coc:2011jh}
Alain Coc, St{\'e}phane Goriely, Yi~Xu, Matthias Saimpert, and Elisabeth
  Vangioni.
\newblock {STANDARD BIG BANG NUCLEOSYNTHESIS UP TO CNO WITH AN IMPROVED
  EXTENDED NUCLEAR NETWORK}.
\newblock {\em Astrophys. J}, 744(2):158, December 2011.

\bibitem{press_1992_aa}
W.~H. {Press}, S.~A. {Teukolsky}, W.~T. {Vetterling}, and B.~P. {Flannery}.
\newblock {\em {Numerical recipes in FORTRAN. The art of scientific
  computing}}, volume 2nd ed.
\newblock {Cambridge: University Press}, 1992.

\bibitem{Kelly:2013jj}
Keegan~J Kelly, Christian Iliadis, Lori Downen, Jordi Jose, and Art Champagne.
\newblock {NUCLEAR MIXING METERS FOR CLASSICAL NOVAE}.
\newblock {\em The Astrophysical Journal {\ldots}}, 777(2):130, October 2013.

\bibitem{MEY13}
B.~S. {Meyer} and M.~J. {Bojazi}.
\newblock {Sensitivity of Nitrogen-15 Production in Explosive Helium Burning to
  Supernova Energies and Reaction Rates and Importance for Low-Density
  Supernova Graphite Grains}.
\newblock In {\em Lunar and Planetary Science Conference}, volume~44 of {\em
  Lunar and Planetary Inst. Technical Report}, page 3006, March 2013.

\bibitem{Iocco:2007jv}
F~Iocco, G~Mangano, G~Miele, O~Pisanti, and P~Serpico.
\newblock {Path to metallicity: Synthesis of CNO elements in standard BBN}.
\newblock {\em Physical Review D}, 75(8):087304, April 2007.

\bibitem{ekstrom_2008_aa}
S.~{Ekstr{\"o}m}, G.~{Meynet}, C.~{Chiappini}, R.~{Hirschi}, and A.~{Maeder}.
\newblock {Effects of rotation on the evolution of primordial stars}.
\newblock {\em \aap}, 489:685--698, October 2008.

\end{thebibliography}

\end{document}